\documentclass[aps,prx,longbibliography,reprint]{revtex4-2}
\usepackage[intlimits]{amsmath}
\usepackage{amssymb}
\usepackage{bbm}
\usepackage[]{units}
\usepackage{xcolor}
\usepackage{hyperref}
\usepackage{braket}
\usepackage{dsfont}
\usepackage{mathdots}
\usepackage{natbib}
\usepackage[T1]{fontenc}
\usepackage{caption}
\usepackage{subcaption}
\usepackage{ragged2e}
\usepackage[textwidth=1.5cm]{todonotes}
\usepackage{upgreek}
\usepackage{placeins}
\newcommand{\bv}{\left( \begin{array}{c}}
\newcommand{\ev}{\end{array} \right)}

\newcommand{\usw}{$U_\mathrm{SWAP}\,$}

\newcommand{\dez}{\Delta E_\mathrm{Z}}
\newcommand{\fosc}{f_\mathrm{osc}}
\newcommand{\dt}{\mathrm{d}t}

\begin{document}
\clearpage
%%%%%%%%%%%%%%%%%%%%%%%%%%%%%%%%%%%%%%%%%
\title{Pulse Shaping for Ultra--Fast Adiabatic Quantum Gates}
\author{İlker Polat$^{1,2}$}
\author{Ramon W.J. Overwater$^{1,2}$}
\author{Maximilian Rimbach-Russ$^{1,3}$}
\author{Fabio Sebastiano$^{1,2}$}
\address{$^{1}$QuTech, Delft University of Technology, Lorentzweg 1, 2628 CJ Delft, The Netherlands}
\address{$^{2}$Department of Quantum and Computer Engineering, Delft University of Technology, 2628 CJ Delft, The Netherlands}
\address{$^{3}$Kavli Institute of Nanoscience, Delft University of Technology, Lorentzweg 1, 2628 CJ Delft, The Netherlands}

\begin{abstract}
A fundamental challenge in quantum computing is to increase the number of operations within the qubit coherence time. While this can be achieved by decreasing the gate duration, the use of shorter signals increases their bandwidth and can cause leakage into energetically separated states.
A common method to suppress leakage for short pulses is the Derivative Removal by Adiabatic Gate (DRAG) method, which however, relies on IQ modulation of radio-frequency (RF) signals, thus cannot be applied to the baseband signals, e.g., for semiconductor spin qubits.
This paper proposes a novel technique, Delayed Leakage Reduction (DLR), that suppresses leakage at targeted frequencies even for baseband control by using time-delayed repetitions of the control signal to enable rapid, high-fidelity operations. We apply DLR on the adiabatic CZ gate between two spin qubits and achieve fidelities exceeding $99.9$\% within \unit[9.4]{ns} for a resonance frequency difference of only \unit[100]{MHz}. Towards the experimental realization of the proposed control method, we also assess the impact on the fidelity of the sampling rate of the electronic hardware generating the control pulse, thus setting the minimum hardware requirements for any experimental demonstration.
\end{abstract}
\maketitle

%111111111111111111111111111111111111111
\section{Introduction}
\label{sec:intro}

Semiconductor spin qubits~\cite{DiVincenzo_1998,burkardSemiconductorSpinQubits2023} are among the leading candidates for realizing scalable quantum computers thanks to their compatibility with classical CMOS fabrication technology~\cite{vandersypenInterfacingSpinQubits2017,zwerverQubitsMadeAdvanced2022,steinacker300MmFoundry2024a,neyensProbingSingleElectrons2024,george12spinqubitArraysFabricated2024} and their high-temperature operation~\cite{Vandersypen2017,petitDesignIntegrationSinglequbit2022, Camenzind2022,undsethHotterEasierUnexpected2023, Huang2024}. 
While their potential for quantum computing has been demonstrated in various experiments ~\cite{morelloSingleshotReadoutElectron2010,mauneCoherentSinglettripletOscillations2012,plaSingleatomElectronSpin2012,yangOrbitalValleyState2012,kawakamiElectricalControlLonglived2014,muhonenStoringQuantumInformation2014,veldhorstAddressableQuantumDot2014,engIsotopicallyEnhancedTriplequantumdot2015,reedReducedSensitivityCharge2016,lawrieSpinRelaxationBenchmarks2020,stanoReviewPerformanceMetrics2022,philipsUniversalControlSixqubit2022,xueQuantumLogicSpin2022a,madzikPrecisionTomographyThreequbit2022a,wang_hopping_2024,noiriFastUniversalQuantum2022a,takedaQuantumErrorCorrection2022, millsTwoqubitSiliconQuantum2022,tanttuAssessmentErrorsHighfidelity2024}, 
 high-fidelity single and two-qubit gates are essential to achieve a fully operational quantum computer~\cite{raussendorfFaultTolerantQuantumComputation2007,hetenyiTailoringQuantumError2024}.

High-fidelity two-qubit gates between spin qubits are realized using the electrically tunable exchange coupling $J$, which arises from the overlap of the wavefunctions~\cite{burkardSemiconductorSpinQubits2023} and can be precisely controlled by electric signals applied to the gate electrodes. The exchange interaction can be used to implement the maximally entangling $\sqrt{\mathrm{SWAP}}$, controlled X (CNOT), and the controlled Z (CZ) quantum gates depending on the single-qubit Zeeman energies. The experimental implementation of $\sqrt{\mathrm{SWAP}}$ operations is often challenging due to the stringent requirements on the resonance frequency difference $\dez$, $J \gg \dez$~\cite{burkardPhysicalOptimizationQuantum1999,meunierEfficientControlledphaseGate2011} and rapid pulse rise times~\cite{petitDesignIntegrationSinglequbit2022}. The highest reported fidelities were realized using the CZ gate~\cite{stanoReviewPerformanceMetrics2022}, that is implemented by varying the exchange interaction adiabatically with respect to $\dez$ to suppress qubit excitations~\cite{meunierEfficientControlledphaseGate2011,Rimbach-Russ_2023}.
In practice, the exchange interaction $J$ is controlled by tuning the tunnel coupling between the two qubits via barrier gate electrodes, and $\dez$ is realized either by local differences in the g-factor~\cite{kloeffelProspectsSpinBasedQuantum2013,scappucciGermaniumQuantumInformation2020} or designed via anisotropic magnetic field via micromagnets~\cite{tokuraCoherentSingleElectron2006,pioro-ladriereElectricallyDrivenSingleelectron2008,philipsUniversalControlSixqubit2022}.

The CZ gate fidelity depends on the duration, resonance frequency splitting, decoherence, and shape of the control pulse ~\cite{xueQuantumLogicSpin2022a,Rimbach-Russ_2023}. While fast gate times reduce decoherence, nonadiabatic pulses can cause the fidelity to rapidly oscillate as a function of the pulse duration due to interference of the excitations. Therefore, high-fidelity CZ-gate implementations require precise timings commensurate with the frequency $\sqrt{\dez^2+J^2}$~\cite{burkardPhysicalOptimizationQuantum1999,russHighfidelityQuantumGates2018,heinzCrosstalkAnalysisSinglequbit2021,petitDesignIntegrationSinglequbit2022,Rimbach-Russ_2023}.
In contrast, adiabatic pulses gradually vary the Hamiltonian parameters such that the system can follow the changes, resulting only in a phase accumulation during the evolution. Recent experiments employing adiabatic pulses have demonstrated two-qubit-gate  fidelity beyond $99$\% ~\cite{xueQuantumLogicSpin2022a,madzikPrecisionTomographyThreequbit2022a,noiriFastUniversalQuantum2022a,millsTwoqubitSiliconQuantum2022,tanttuAssessmentErrorsHighfidelity2024,stanoReviewPerformanceMetrics2022}.

The standard approach for mitigating errors in gate operations follows the optimal control theory~\cite{glaserTrainingSchrodingerCat2015}, in which
optimal pulse are derived using strategies commonly referred to as shortcuts to adiabaticity~\cite{bergmannCoherentPopulationTransfer1998,F.Motzoi_2009,berryTransitionlessQuantumDriving2009,degrandiAdiabaticPerturbationTheory2010,ivakhnenkoNonadiabaticLandauZener2023,vuthaSimpleApproachLandauZener2010,chernovaOptimizingStateTransfer2024, stepanenkoTimeoptimalTransferQuantum2025,motzoiOptimalControlMethods2011,chenLewisRiesenfeldInvariantsTransitionless2011,ribeiroSystematicMagnusBasedApproach2017,selsMinimizingIrreversibleLosses2017,theisCounteractingSystemsDiabaticities2018,banFastLongrangeCharge2018,takahashiHamiltonianEngineeringAdiabatic2019,banSpinEntangledState2019,guery-odelinShortcutsAdiabaticityConcepts2019,setiawanAnalyticDesignAccelerated2021,zhuangNoiseresistantLandauZenerSweeps2022,takahashiDynamicalInvariantFormalism2022,glasbrennerLandauZenerFormula2023,Rimbach-Russ_2023,dengisAcceleratedCreationNOON2025,romeroOptimizingEdgestateTransfer2024, liuAcceleratedAdiabaticPassage2024,xuImprovingCoherentPopulation2019,fehseGeneralizedFastQuasiadiabatic2023,limaSuperadiabaticLandauZenerTransitions2024,richermeExperimentalPerformanceQuantum2013, rolandQuantumSearchLocal2002,martinez-garaotFastQuasiadiabaticDynamics2015,chenSpeedingQuantumAdiabatic2022,fernandez-fernandezQuantumControlHole2022, meinersenQuantumGeometricProtocols2024,meinersenUnifyingAdiabaticStatetransfer2025}.
For instance, the use of window functions for the pulse shape can improve the fidelity limited by non-adiabatic errors, as leakage control errors are related to the spectrum of the control pulse~\cite{martinisFastAdiabaticQubit2014,Rimbach-Russ_2023}. A stronger suppression of non-adiabatic dynamics can be realized by using IQ modulation, such as Derivative Removal by Adiabatic Gate (DRAG)~\cite{F.Motzoi_2009} or oscillating signals around zero~\cite{9583615}. However, since the exchange interaction for CZ gates in spin qubits is always a positive-valued baseband signal, these methods are not applicable.

In this work, we demonstrate the relation between the gate fidelity of an adiabatic CZ gate and the spectrum of the control pulse. Using the gained insight, we introduce a simple time-domain pulse-shaping strategy, Delayed Leakage Reduction (DLR), that suppresses signal leakage to other energy levels using only baseband pulses. Our method combines two identical control signals with a time delay between them, specifically engineered to suppress leakage at targeted frequencies that could otherwise degrade the fidelity. Furthermore, we use our method to achieve a gate time of only \unit[9.4]{ns} for $\dez=$ \unit[100]{MHz}  with fidelity $>99.9$\%, thus approaching the maximum possible gate speed. Additionally, we analyze the impact of a limited sampling rate for the adiabatic control signal on the fidelity to assess the feasibility of our proposed method in a practical experimental setup. By determining the minimum sampling rate required for adiabatic pulses, we remarkably find that the proposed DLR method only marginally increases the required sampling rate compared to traditional window methods.

In Section \ref{sec:method}, we start by introducing the methods necessary to compute the fidelity and excitation rate of the evolution based on the spectrum of the control signal and $\dez$. Section \ref{sec:dlr} introduces our new pulse shaping method, DLR, that suppresses the leakage to undesired states. Later, in Section \ref{sec:nyquist}, we show derive and estimate the minimum sampling rate for adiabatic signals. Finally, in Section \ref{sec:conclusion}, we summarize our results present prospective future work.

\section{Results}
\label{sec:method}
\subsection{Analysis of CPHASE gates between spin qubits}
The controlled Z (CZ) gate is obtained as a special case of the controlled phase (CPHASE) gate, which naturally arises from the exchange interaction between spins. More precisely, the CZ operation can be achieved by composing a CPHASE gate with phase $\pi$ and single-qubit Z operations
\begin{align}
\mathrm{CZ}=\mathrm{Z_1}(-\pi/2)\:\mathrm{Z_2}(-\pi/2)\:\mathrm{U_{CPHASE}}.
\label{eq:czfromcphase}
\end{align}
This decomposition highlights that the two gates differ only by local, single-qubit phases, which often can be implemented virtually and do not affect entanglement.

The Hamiltonian of two interacting spin qubits via the exchange interaction can be written in the standard computational basis $\{\ket{00},\ket{01},\ket{10},\ket{11}\}$ as~\cite{meunierEfficientControlledphaseGate2011}
\begin{align}
H_\mathrm{exc} = \begin{pmatrix}
    E_z & 0         & 0         & 0 \\
    0   & (-\dez-J)/2     & J/2         & 0 \\
    0   & J/2       & (\dez-J)/2      & 0 \\
    0   & 0         & 0         & -E_z
\end{pmatrix},
\label{eq:hexec}
\end{align}
where $E_z=(f_1+f_2)/2$ and $\dez=f_1-f_2$ are the average and difference of the qubit resonance frequencies $f_1$ and $f_2$, respectively. $J$ represents the exchange interaction that can be controlled by detuning or tunneling the coupling of the spins. The time-evolution of this Hamiltonian directly implements $\sqrt{\mathrm{SWAP}}$ and $\mathrm{CPHASE}$ operations depending on the choice of $\dez$ and the waveform used for $J(t)$.
To facilitate a more transparent analysis of the system dynamics, we decompose the total Hamiltonian $H_{\mathrm{exc}}$ into three mutually commuting components
\begin{align}
H_{\mathrm{RF}} &= \begin{pmatrix}
    E_z & 0         & 0         & 0 \\
    0   & 0     & 0         & 0 \\
    0   & 0       & 0      & 0 \\
    0   & 0         & 0         & -E_z
\end{pmatrix},\\
H_{\mathrm{CPHASE}} &= \begin{pmatrix}
    0 & 0         & 0         & 0 \\
    0   & -J/2     & 0         & 0 \\
    0   & 0       & -J/2      & 0 \\
    0   & 0         & 0         & 0
\end{pmatrix},\\
H_{\mathrm{SWAP}} &= \begin{pmatrix}
    0 & 0         & 0         & 0 \\
    0   & (-\dez)/2     & J/2         & 0 \\
    0   & J/2       & (\dez)/2      & 0 \\
    0   & 0         & 0         & 0
\end{pmatrix}.
\label{eq:hswap}
\end{align}
By separating the Hamiltonian in this manner, we gain direct access to the individual unitary evolutions, enabling a clearer understanding of how each term contributes to the overall gate operation and how undesired dynamics can be suppressed.
Since all three components commute, the total unitary corresponding to $H_{exc}$ is given by the product of the time-evolution of each individual Hamiltonian. The Hamiltonians $H_{\mathrm{CPHASE}}$ and $H_{\mathrm{SWAP}}$ can be used to implement $\mathrm{CPHASE}$ and $\mathrm{SWAP}$ operations respectively, and $H_{\mathrm{RF}}$ implements the same single-qubit Z gate on both qubits simultaneously. As these joint single-qubit operations can be calibrated out by tracking the time and the resonance frequency or compensated by individual follow-up gates, we will ignore them for the remainder of this article. 

To achieve high fidelity $\mathrm{CPHASE}$ ($\mathrm{SWAP}$) operations, the effect of the other Hamiltonian, i.e., $H_\mathrm{SWAP}$ ($H_\mathrm{CPHASE}$) must be suppressed, i.e., its resulting unitary should be as close as possible to the identity operator ($I$). 
%Below, we investigate multiple approaches for achieving high-fidelity $\mathrm{CZ}$ operations by suppressing the oscillatory dynamics associated with the $\mathrm{SWAP}$ component.
For the CPHASE operation $H_{\mathrm{CPHASE}}$, the corresponding time-evolution operator is given by
\begin{align}
U_{\mathrm{CPHASE}} &= \begin{pmatrix}
    1 & 0         & 0         & 0 \\
    0   & e^{i  \, \pi\: \int_0^t J(t) dt}     & 0         & 0 \\
    0   & 0       & e^{i  \,\int_0^t J(t) dt}      & 0 \\
    0   & 0         & 0         & 1
\end{pmatrix}.
\label{eq:ucphase}
\end{align}
%Since $H_{\mathrm{CPHASE}}$ is diagonal, $J$ can be replaced with a time-dependent $J(t)$ below. 
By choosing $\int 2\pi \cdot J(t) \, \mathrm{d}t = \pi$, the diagonal terms are $(1, i , i , 1)$, so that a CZ gate can be implemented using Eq.~\ref{eq:czfromcphase}.

\begin{figure*}[htbp]
    %\centering
    \justifying
    \begin{subfigure}[b]{0.30\textwidth}
        \centering
        \includegraphics[width=\textwidth]{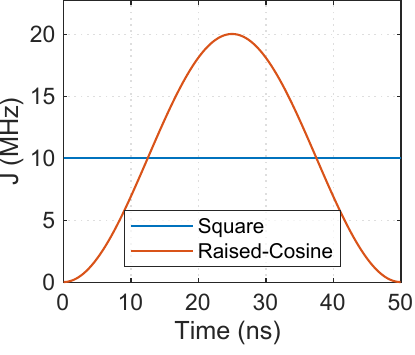}
        \caption{}
        \label{fig:1a}
    \end{subfigure}
    \hfill
    \begin{subfigure}[b]{0.30\textwidth}
        \centering
        \includegraphics[width=\textwidth]{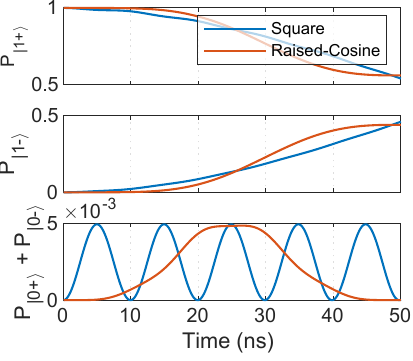}
        \caption{}
        \label{fig:1b}
    \end{subfigure}
    \hfill
    \begin{subfigure}[b]{0.30\textwidth}
        \centering
        \includegraphics[width=\textwidth]{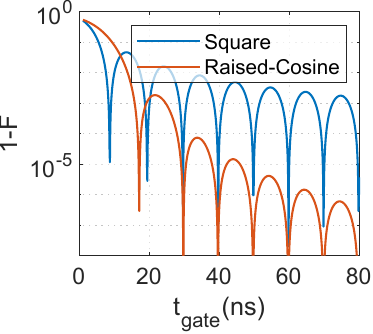}
        \caption{}
        \label{fig:1c}
    \end{subfigure} 
    \caption{Simulated $\mathrm{CPHASE}$ operation on $H_\mathrm{J}$ with  $\dez=100\,\mathrm{MHz}$ for  square (blue) and raised-cosine (red)  $J$ control signals. (a) Time evolution of $J(t)$, with duration $t_\text{gate}=$50 ns and amplitude tuned for $\int 2\pi \cdot J(t) \, \mathrm{d}t = \pi$ (b) State probabilities of $P_{|1+\rangle}$,$P_{|1-\rangle}$ and $P_{|0+\rangle}+P_{|0-\rangle}$ during the operation (c) Fidelity of the CZ operation when sweeping the gate duration ($t_\text{gate}$) for both square (blue) and raised-cosine (red) shaped $J$ signals,  assuming perfect single-qubit phase compensation. For each point, the amplitude of the signal is optimized for the highest fidelity.}
\label{fig:state_evolution}
\end{figure*}

A high-fidelity $\mathrm{CPHASE}$ gate can be realized if the time evolution of the $\mathrm{SWAP}$ Hamiltonian \usw is equal to the identity operator up to single-qubit phase errors.
Before addressing the general case of a time-varying $J$, we briefly analyze the case for constant $J$
\begin{align}
    U_\text{SWAP}=&\frac{\mathds{1}+ZZ}{2}+\cos(\pi \fosc t)\frac{\mathds{1}-ZZ}{2} -i \sin(\pi \fosc t)\nonumber\\ &\times\bigg[\frac{\dez}{2\fosc}(IZ-ZI)
    +\frac{J}{2\fosc}(XX+YY)\bigg],
    \label{eq:uswap}
    %     U_\text{SWAP}=\cos(\pi \fosc t)\mathds{1}_2 &-i \frac{2H_{\mathrm{SWAP}}}{\fosc}\sin(\pi \fosc t),
    % \label{eq:uswap}
\end{align}
where $PQ=P\otimes Q$ with $P,Q=I,X,Y,Z$ being two-qubit Pauli operators and $\fosc=\sqrt{\dez^2+J^2}$. 
Residual single-qubit phase errors ($ZI$ and $IZ$) can be compensated in this case with cascaded virtual single-qubit $\mathrm{Z}(\theta)$ gates~\cite{xueQuantumLogicSpin2022a}. However, the presence of the non-diagonal terms ($XX$ and $YY$) leads to leakage to other energy levels and prevents the adoption of simple phase compensations. 
To suppress the non-diagonal terms, three commonly employed strategies are available. Firstly, by choosing $J\ll \dez$, the amplitude of the non-diagonal terms in Eq.~\eqref{eq:uswap} will be $J/f_{\mathrm{osc}} \approx J/\dez\ll 1$. This reduces the oscillation amplitude, enabling the desired fidelity, but requires a longer gate time to accumulate sufficient phase in Eq.~\eqref{eq:ucphase} due to the smaller value of $J$ or ask for a larger frequency separation $\dez$, which can be challenging to achieve in practice. Secondly, the gate time can be chosen as an integer multiple of $\fosc$, which leads the $\sin(\pi \fosc t)$ to be zero at the end of the evolution. However, this method, called synchronization, requires extremely precise pulse timing, especially for faster gates~\cite{Rimbach-Russ_2023}. Lastly, pulse shapes for $J(t)$ that follow classical window functions can be used~\cite{martinisFastAdiabaticQubit2014,Rimbach-Russ_2023}. This method enforces slow changes of the eigenvectors of the system, thus resulting in a higher likelihood of ending in the initial state. We will examine this technique in more detail in the next subsection, where the assumption of a time-constant $J$ is dropped.

\subsection{Relating non-adiabatic errors to spectral properties}
\label{sec:relating_fidelity_to_spectrum}
In Fig.~\ref{fig:state_evolution}, we compare the pulses and the resulting evolutions for the two commonly adapted strategies, synchronization and the usage of window-functions for pulse shaping. 
In an ideal $\mathrm{CPHASE}$ operation, only the transition from $P_{|1+\rangle}$ to $P_{|1-\rangle}$ should occur, with both populations reaching 1/2 at the end of the gate. However, the presence of the undesired $\mathrm{SWAP}$ components induces additional oscillations in the residual population $P_{|0+\rangle} + P_{|0-\rangle}$, and may prevent $P_{|1+\rangle}$ and $P_{|1-\rangle}$ from reaching the aimed 1/2. 
Although $P_{|1+\rangle}$ and $P_{|1-\rangle}$ states can be corrected to 1/2 via single qubit $Z$ calibration (see App.~\ref{appx:relation_fidelity_and_spectrum}), compensating the residual state populations $P_{|0+\rangle}+P_{|0-\rangle}$ requires the implementation of real single qubit operations. 

Furthermore, choosing a smaller $J/\dez$ ratio will reduce the amplitude of the oscillations in the residual state at the cost of longer gates for fixed $\dez$. As an alternative, since the residual state evolution is periodic with frequency $f_\mathrm{osc}$ for the square pulse, a perfect gate can be achieved by synchronizing the gate time with the gate oscillations~\cite{burkardPhysicalOptimizationQuantum1999,russHighfidelityQuantumGates2018,heinzCrosstalkAnalysisSinglequbit2021,Rimbach-Russ_2023}. However, using a perfect square pulse requires wide bandwidth and precise timing, as highlighted in Fig.~\ref{fig:1c}. In contrast, a smooth adiabatic pulse, e.g., a raised-cosine pulse, greatly decreases the residual state error (see Fig.~\ref{fig:1b}) during the evolution, thus allowing for faster gate times without strict requirements on the timing of the control signal. As a result, adiabatic pulses longer than \unit[15]{ns} can achieve a higher gate fidelity than the square pulses (see Fig.~\ref{fig:1c}). However, shorter high-fidelity gates cannot be implemented straightforwardly using a raised-cosine (or other window function) pulse~\cite{Rimbach-Russ_2023}.

To better understand the characteristic dynamic of the adiabatic evolution~\cite{Rimbach-Russ_2023}, we make use of an approximate analytical expression of the non-adiabatic error (see appendix~\ref{app:errors})
\begin{align}
\epsilon=|\text{tr}(\mathcal{E} \ket{01}\bra{10})|^2 \approx \left|
\pi
\int_0^{t_\text{gate}} J(t) e^{i  2\pi ( \alpha_1(t)-\alpha_2(t))} dt \right|^2
    \label{eq:sparseEvolution}.
\end{align}
where $\alpha_{1}$ and $\alpha_{2}$ are the accumulated phase of the eigenstates $\ket{01}$ and $\ket{10}$, i.e., $\alpha_{1,2}(t)= \pm\int_0^t  \fosc(t')/2\, dt' $, and $t_\text{gate}$ is the gate duration. This expression can be interpreted as the control signal interacting with the frequency of one state ($e^{i2\pi \alpha_1}$) and being observed in a rotating frame of the other qubit state ($e^{-i2\pi\alpha_2}$). 
To make this formulation more practical for spectral analysis, we rewrite the phase difference in the exponent by introducing and subtracting $\dez$. This isolates a term that oscillates at $\dez$ as 
\begin{align}
    \epsilon \approx \left| \pi \int _{0}^{t_\text{gate}} J(t) e^{i  2\pi \,\int_0^{t}( \fosc(t') - \dez )\mathrm{d}t'} e^{i  2\pi t \, \dez} \dt\right|^2.
    \label{eq:sparseEvolution_2}
\end{align}

For $J \ll \dez$, the difference $\fosc(t)-\dez$ becomes negligible and the integral simplifies. 
Under this approximation, the error can be directly related to the magnitude of the Fourier spectrum of $J(t)$ at frequency $\dez$

\begin{align}
	\epsilon\approx |\pi \, S(-\dez)|^2.
 \label{eq:fourier_error_rate}
\end{align}
where 
\begin{align}
	S(f)&=\int_0^{t_\text{gate}} J(t) e^{-i  2\pi t  f}dt.
\end{align}
is the Fourier transform of $J(t)$.

The average gate infidelity of the CZ gate is then given by (details and derivation in App.~\ref{appx:relation_fidelity_and_spectrum}) 
\begin{align}
    1-F\approx \frac{2}{5}\left(\epsilon+2\delta\Phi_J^2+\delta\Phi^2+\delta\theta^2\right),
    \label{eq:fidelityAnalytic}
\end{align}
where $\delta\Phi_J=\pi-2\pi\int_0^{t_\text{gate}}J(t)dt$ is the exchange miscalibration, $\delta\Phi= (\theta_1 + \theta_2-\pi)-2\pi E_z t_\text{gate}$ the miscalibration of the rotating frame, $\delta\theta = (\theta_1 - \theta_2)-2\pi\int_0^{t_\text{gate}}\sqrt{\Delta E_z^2+ J^2(t)}dt$ the miscalibrated phase corrections, and $\theta_{1,2}$ is the single-qubit phase gate $Z(-\theta_{1,2})$ for qubit~1~and~2.  

\begin{figure}[tbp]
\centering
\includegraphics[width=.7\columnwidth]{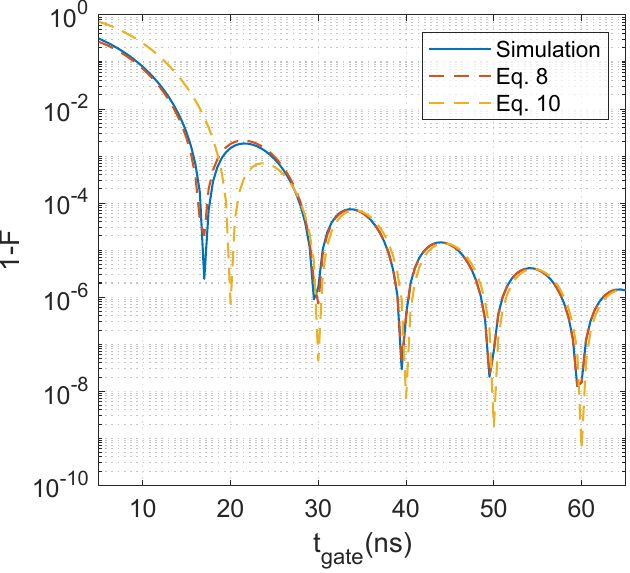}
\caption{The simulated fidelity of the CPHASE gate using a raised-cosine pulse is compared to error rate equations as a function of gate time. The raised-cosine pulse amplitude is optimized for each gate duration to ensure precise phase accumulation, and $\dez$ is $\unit[100]{MHz}$. }
\label{fig:fidelity_functions}
\end{figure}

In Fig~\ref{fig:fidelity_functions}, the fidelity computed using Eq.~\eqref{eq:sparseEvolution} and \eqref{eq:fourier_error_rate} is compared with the simulation results. For different gate durations $t_\text{gate}$, the amplitude of the signal is calibrated such that $\int_0^{t_\text{gate}} 2\pi  J(t) \, \mathrm{d}t = \pi$ is fulfilled. This leads to an increase in $J$ for shorter pulses and the approximation $\fosc(t)-\dez\approx0$ does not to hold anymore. As a consequence, Eq.~\eqref{eq:fourier_error_rate} loses its accuracy for shorter gates, while Eq.~\eqref{eq:sparseEvolution} fits the simulations better. 
This comparison confirms that the analytical model accurately captures the fidelity behavior, especially for longer gate durations.
It is important to note that the spectral analysis of gate fidelity is equally applicable to square pulses.
Such a spectrum exhibits notches at approximately every integer multiple of $1/t_{\mathrm{gate}}$ reflecting the synchronization condition, and high fidelity gates are obtained when one of these notches aligns with frequency $-\dez$.

This spectral perspective reframes the challenge of minimizing gate infidelity as a problem of pulse shaping or spetrcal design. In the next section, we introduce a novel time-domain technique—Delayed Leakage Reduction (DLR)—designed to suppress spectral leakage at critical frequencies.

\subsection{Delayed Leakage Reduction}
\label{sec:dlr}

\begin{figure*}[htbp]
    \subcaptionbox{\label{fig:dlr_example}}{\includegraphics[width=0.3\textwidth]{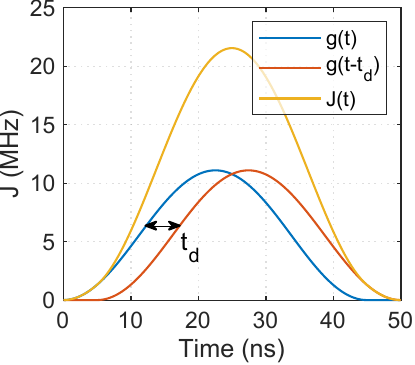}}
    \hfill
    \subcaptionbox{\label{fig:tukey_vs_dlr_fft}}{\includegraphics[width=0.3\textwidth]{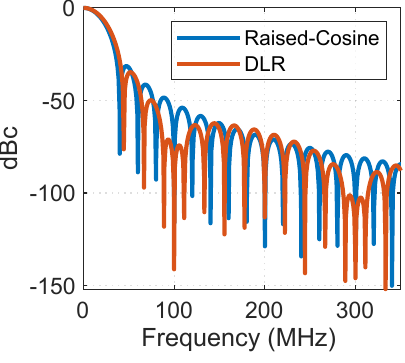}}
    \hfill
    \subcaptionbox{\label{fig:dlr_comparison}}{\includegraphics[width=0.3\textwidth]{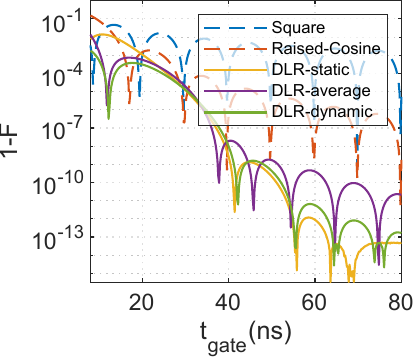}}
    \vfill
    \subcaptionbox{\label{fig:fig_HKT_time}}{\includegraphics[width=0.3\textwidth]{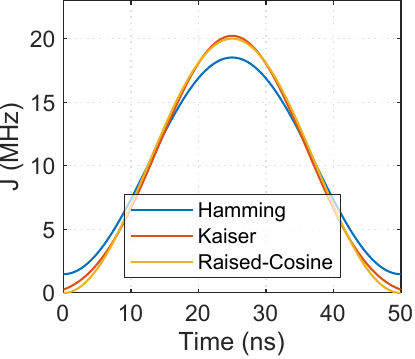}}
    \hfill
    \subcaptionbox{\label{fig:fig_HKT_fft}}{\includegraphics[width=0.3\textwidth]{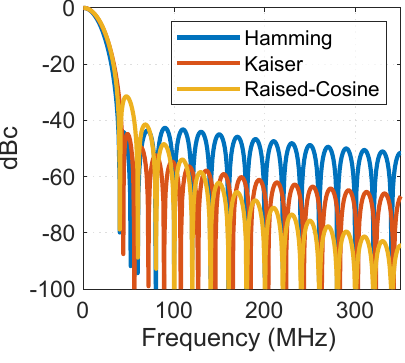}}
    \hfill
    \subcaptionbox{\label{fig:fig_DLR_static_sweep}}{\includegraphics[width=0.3\textwidth]{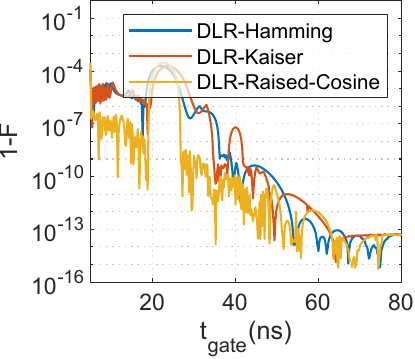}}
    \caption{(a) Example of a \unit[50]{ns} signal generated with DLR. It consists of two raised-cosine pulses, $g(t)$ and $g(t-t_d)$, which have a delay of $t_d$=\unit[5]{ns} between each other to create a notch at $\dez$ =\unit[100]{MHz}. (b) The frequency spectrum of two $J(t)$ signals, a raised-cosine and DLR, with \unit[50]{ns} gate duration. Since DLR consists of two shorter pulses, its main and side lobes are wider. Notches at  integer multiples of $\frac{1}{2t_d}=$\unit[100]{MHz} are clearly visible. (c) Fidelity of the CPHASE operation in a two-qubit system with $\dez$ =\unit[100]{MHz} for different gate durations ($t_\text{gate}$). For every gate time and signal shape, the amplitude of $J(t)$ is tuned for maximum fidelity, according to Eq.\eqref{eq:ucphase}. The DLR-static and DLR-average introduce a spectral notch at fixed frequencies, namely $\unit[100]{MHz}$ and $\frac{1}{2}\left(100 + \max[f_{\mathrm{osc}}(t)]\right)$, respectively. In contrast, the DLR-dynamic continuously adapts the notch frequency throughout the gate operation, following the instantaneous oscillation frequency $f_{\mathrm{osc}}(t)$.  (d) Hamming, Kaiser ( $\beta$ = 6.14)(App.~\ref{app:window_functions}), and raised-cosine window functions with \unit[50]{ns} gate time. (e) Frequency spectrum of the window functions in (d). (f) Infidelity in the CPHASE gate controlled by the DLR-static pulse using each of the window functions in (d). For each different window function, the DLR-static pulse is used to derive the gate fidelity. Among all possible delays generating a notch in the frequency range $80$-\unit[130]{MHz}, the one with the minimum infidelity for a given gate duration has been selected and constitutes a data point in the plot.
    }
    \label{fig:dlr_examples}
\end{figure*}

To minimize leakage at $\dez$ according to Eq.~\eqref{eq:fourier_error_rate}, a delayed replica of the control signal can be combined with the original control signal to cancel out spectral components at the critical frequency $f=-\dez$, thus maximizing the gate fidelity.
If $g(t)$ is the original pulse used to modulate $J$, e.g, the raised-cosine pulse described in the previous section, the Delayed Leakage Reduction (DLR) pulse is defined as
\begin{equation}
g'(t)=g(t)+g(t-t_d)
\end{equation}
where $t_d$ is the delay between two pulses. In the Fourier transform, this leads to a frequency--dependent phase shift of the delayed replica  as

\begin{align} 
	S(f) &= \left [1+e^{-i  2\pi f t_d  }\right ] \int _{0}^{t_\text{gate}} g(t) e^{-i  2\pi t  f}\dt \\
             &= \left [1+e^{-i  2\pi f t_d  }\right ] G(f)
             \label{eq:delay_envelope}.
\end{align}
where $G(f)$ is the Fourier transform of $g(t)$.
By choosing $t_d=\frac{1}{2\, \dez}$, a notch in the spectrum at $S(-\dez)$ is created, thus suppressing the non-adiabatic error. Fig.~\ref{fig:dlr_examples} demonstrates an example of a \unit[50]{ns} DLR--shaped pulse, where $g(t)$ is chosen as a \unit[45]{ns} raised--cosine pulse and $t_d=5$ ns (Fig.~\ref{fig:dlr_example}), thus showing a clear notch at $f_\mathrm{notch}=\frac{1}{2t_d}=100 $ MHz (Fig.~\ref{fig:tukey_vs_dlr_fft}).
In Fig.~\ref{fig:dlr_comparison}, we compare the resulting fidelity of the CPHASE operation to a straightforward DLR implementation (DLR-static) for commonly used window function. While DLR--static outperforms the raised-cosine beyond \unit[30]{ns}, the fidelity of the DLR pulse drops significantly for shorter gate times due $\fosc\approx\dez$ becoming invalid. To overcome such a limitation, more advanced DLR techniques are proposed in the following.

As $J$ changes during the gate operation, so does $\fosc$, especially for shorter gate durations. To cope with this, the DLR--average strategy places the notch in the spectrum at the average of the maximum and minimum value of $\fosc(t)$ during the operation.
However, calculating the required $t_d$ from $\fosc(t)$ is not straightforward since $\fosc(t)$ is a function of $J(t)$, which is also a function of $t_d$.
Therefore, $\fosc(t)$ is first calculated with a raised-cosine pulse for a given gate time and $t_d$ is computed accordingly. While this is not an exact solution, DLR-average can reach lower infidelities with faster gates than the DLR-static strategy (Fig.~\ref{fig:dlr_comparison}). 

The last method method DLR--dynamic uses time-dependent $t_d(t)$ to follow $\fosc(t)$. Since in this method $J(t)$ and $\fosc(t)$ are recursive as for DLR--average, $\fosc(t)$ is first calculated with a simple Raised-Cosine pulse. To preserve the gate time, we redefine $g(t)$ as
\begin{align}
g(t,t_\text{gate}) = \left\{
        \begin{array}{ll}
             0, & t < 0 \\
            \frac{1}{2} \left(1 - \cos\left(\frac{2 \pi t}{t_\text{gate}}\right)\right), & 0 \leq t \leq t_\text{gate} \\
            0, & t > t_\text{gate}
        \end{array}
    \right.
\end{align}
Then the pulse for $J(t)$ is given by $g(t,t_\text{gate}-t_d(t))+g(t-t_d(t), t_\text{gate}-t_d(t))$, where $t_\text{gate}$ is the required gate time. As shown in Fig.~\ref{fig:dlr_comparison}, using the DLR-dynamic (purple), after a gate time of  \unit[9.4]{ns}, the infidelity never exceeds $8.9 \times 10^{-4}$.

Although those strategies can improve the fidelity below \unit[15]{ns} with respect to DLR-static, the appearance of a infidelity hump between approximately \unit[20-25]{ns} (see Fig.~\ref{fig:dlr_comparison}) does not change.
To explain this, we find that for a gate time of \unit[20-25]{ns}, the lowest-frequency notch of the DLR-pulse spectrum is only at \unit[80-100]{MHz}. Since the slope of the main lobe is steeper than the other lobes, the notch introduced by the DLR-envelope cannot properly attenuate the spectrum.

Although the control signal discussed so far has been constructed only by adding a delayed replica, it is also possible to use multiple delayed replicas with different amplitudes and delays to customize the spectrum for specific applications, e.g., by reducing the leakage to various energy levels or for stronger suppression. 

\begin{figure*}[htbp]
    \subcaptionbox{\label{fig:fig_drag_re_im}}{\includegraphics[width=0.32\textwidth]{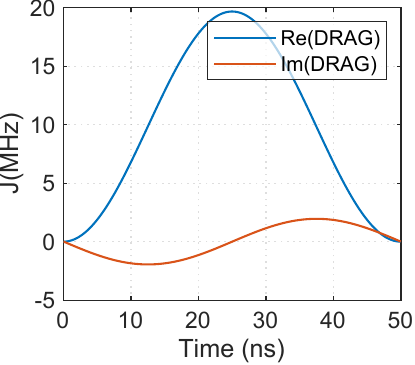}}
    \hfill
    \subcaptionbox{\label{fig:dlr_vs_drag_fft}}{\includegraphics[width=0.32\textwidth]{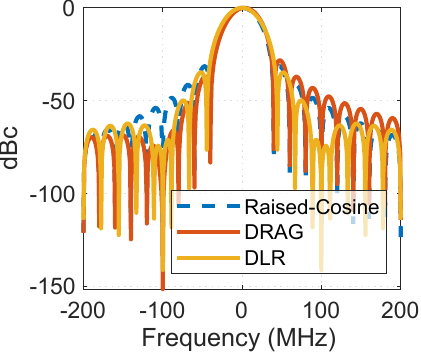}}
    \hfill
    \subcaptionbox{\label{fig:drag_dlr_comparision}}{\includegraphics[width=0.3\textwidth]{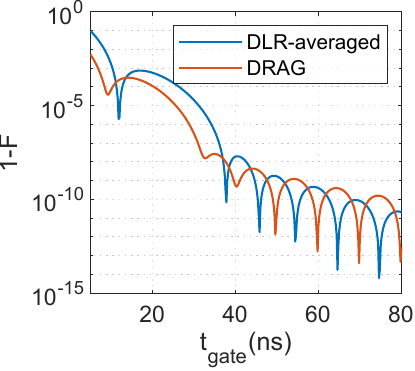}}
    \caption{(a) Example of $J$ pulse generated with DRAG where $\beta={1}/{(2\pi \dez)}$, with $\dez=100$ MHz. (b) Comparison of the DRAG and DLR frequency spectrum with $50$ns $t_\text{gate}$ and notch at \unit[100]{MHz}. (c) Infidelity of DLR and DRAG signals for different gate durations ($t_\text{gate}$). The signal amplitude is optimized for each $t_\text{gate}$ to achieve accurate $\pi$ phase accumulation.}
    \label{fig:dlr_vs_drag}
\end{figure*}

\subsection{Numerical optimization for other windows with DLR}

Adopting a more generic pulse for the $J$-control signal, such as a window function, may become analytically challenging, because, for instance,  window functions with non-zero initial or final value may introduce discontinuities during the DLR pulse due to the delayed replica. Such discontinuities can cause a rapid change in $\fosc$ during the operation, thus invalidating Eq.~\ref{eq:sparseEvolution_2} \cite{Rimbach-Russ_2023}.
However, the optimal delay $t_d$ for a DLR-static pulse can be numerically identified for a given gate time and window shape by directly simulating the fidelity performance of each sample waveform.
In Fig.~\ref{fig:fig_DLR_static_sweep}, the best fidelity for the given $t_\text{gate}$ is shown for Hamming, Kaiser, and the raised-cosine. A hump similar to the one observed in Fig.~\ref{fig:dlr_comparison} can be clearly seen around \unit[20-25]{ns} for the same reason. Due to the different shapes of the main lobe for the three different functions, each pulse has a different hump over the different values of $t_\text{gate}$.
The noisy behavior in those results is attributed to numerical noise, e.g., a too large step size in the $t_d$ sweep.

The results demonstrate that high-fidelity operations exceeding 99.99\% are achievable across a broad range of gate times, provided that the notch frequency is appropriately tuned. The error range of highly tuned $t_{d}$ for each window can be found in Fig.~\ref{fig:DLR_static}d-f.

\subsection{Comparing DLR and DRAG}

The Derivative Removal by Adiabatic Gate (DRAG) protocol is one of the most common quantum control methods used to reduce leakage to higher energy levels. This is achieved by adding correction terms to the control signals to mitigate the leakage to unwanted levels. The first-order DRAG pulse is constructed  by adding the derivative of the original pulse, scaled by a specific coefficient, to the original signal
\begin{align}
	J(t)=g(t)+i \beta\, \frac{d}{dt}g(t),
\end{align}
where $\beta$ is a scaling coefficient. The resulting spectrum is approximately
\begin{align}
	S(f)&=G(f)+i2\pi f(i \beta \,G(f))\\
            &=G(f)[1-2\pi f\beta].
\end{align}
where $G(f)$ is the spectrum of $g(t)$.

Setting $\beta = -\frac{1}{2\pi\,\dez}$ ensures $S(-\dez) = 0$, minimizing leakage at the oscillation frequency. In Fig.~\ref{fig:dlr_vs_drag_fft}, the frequency spectrum of DRAG and DLR are compared. While the spectrum of the DRAG technique is suppressed at a single side, the DLR technique achieves suppression at both sides.

In Fig.~\ref{fig:drag_dlr_comparision}, the DLR--average and DRAG techniques applied to the raised-cosine pulse are compared as a function of the gate duration. 
For a fair comparison, $\beta$ is also chosen to create the notch at the average of the maximum and minimum values of $\fosc$. 
The plots for both methods are similar, except that DLR follows the DRAG with a small shift in gate duration. A closer inspection indicates that the value of the shift is approximately the time delay $t_d$ of our DLR protocol. 
As we have discussed in Sec.\ref{sec:dlr}, DLR expands the spectrum due to the use of two pulses that are $t_d$ shorter than the original pulse, thus expanding the spectrum of the resulting pulse. This expansion explains why the DRAG pulse outperforms DLR slightly.

However, from a practical standpoint, implementing a DRAG pulse requires access to the complex domain, e.g., by adopting IQ bandpass modulation as usually done for transmon qubits \cite{motzoiSimplePulsesElimination2009,PhysRevA.88.052330}, which is infeasible for baseband pulses such as the exchange interaction for the CPHASE gate implementations~\cite{xueQuantumLogicSpin2022a} and similar situations with limited control. Additionally, DLR features an always symmetric pulse shape. This symmetry could be particularly advantageous in scenarios such as simultaneous spin-qubit driving \cite{wu2025simultaneoushighfidelitysinglequbitgates}, where control pulses must selectively address a target qubit without inducing leakage or excitations in neighboring qubits. Consequently, although DLR performs slightly worse than DRAG, its straightforward time-domain implementation and symmetric spectrum make it more suitable for specific applications.

\begin{figure}[!b]
    \centering
    \includegraphics[width=.6\columnwidth]{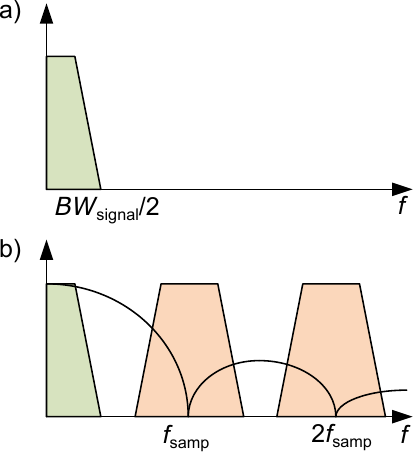}
    \caption{Demonstration of the Nyquist sampling criterion in the frequency domain. (a) The spectrum of an arbitrary baseband signal, with its bandwidth indicated. (b) The spectrum after sampling, showing the replication of the signal at integer multiples of the sampling frequency. The sinc-shape envelope is introduced by zero-order hold sampling.}
    \label{fig:sampling_ilu}
\end{figure}

\subsection{Nyquist criterion on adiabatic sampling}
\label{sec:nyquist}
In practice, adiabatic pulses for $J$ will be generated through the use of signal generators, such as arbitrary wave generators (AWGs), that reproduce a time-sampled version of $J(t)$. However, sampling at a rate lower than the Nyquist rate can increase leakage or excitations due to the appearance of unwanted oscillations at frequencies $\fosc$. While choosing a sampling frequency much higher than the minimal requirement could be a safe option, it significantly increases the power and cost of the electronic interface, thus limiting the scalability and the feasibility of such a solution. Therefore, determining the minimum sampling rate required to preserve adiabaticity of the evolution is of utmost importance.

We start by investigating the effect of finite sampling on the spectrum $S(f)$ of the DLR pulse.
After sampling a signal in the time domain, aliases of the signal are created at integer multiples of the sampling frequency, as shown in Fig.~\ref{fig:sampling_ilu}. Furthermore, assuming that the analog electronics reconstructing the pulse adopts a zero-order hold, the $\mathrm{sinc}(f)$ envelope modulate the entire spectrum. The frequency spectrum of $S(f)$, sampled with the sampling frequency $f_\mathrm{samp}$ can then be written as \cite{Quinquis_2008,Pelgrom_2022}.
\begin{align}
    S_\text{samp}(f)=  \frac{\sin(\pi f/f_\mathrm{samp})}{\pi  f/f_\mathrm{samp}}\sum_{k=-\infty}^{\infty}  S(f-f_\mathrm{samp}\cdot k ).
\end{align}

\begin{figure}[!t]
    \centering
    \includegraphics[width=.8\columnwidth]{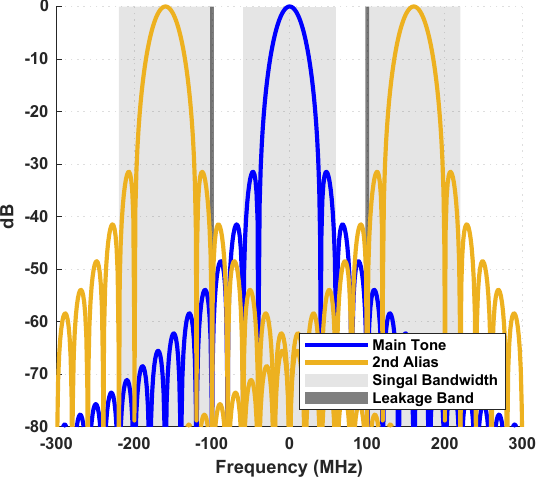}
    \caption{An example of how Eq.\eqref{eq:sampling_cri} determines the minimum sampling frequency for a raised-cosine pulse with duration of \unit[50]{ns}. With an infidelity threshold $F_{th}=0.01\,\%$, $BW_\mathrm{signal}$ is defined asthe frequency at which the spectral power drops by $10\cdot log_{10}(F_{th})=\unit[-40]{dB}$ with respect to the main lobe, leading to  $BW_\mathrm{signal}=\unit[120]{MHz}$. Since higher-order aliases are further from the baseband, it is sufficient to examine only the second alias. 
    With Eq.\eqref{eq:sampling_cri}, the minimum sampling frequency of \unit[160]{MHz} ensures that the second alias lies just beyond the oscillation frequency, thereby avoiding spectral overlap. Sampling at frequencies above this threshold further separates the alias from \unit[160]{MHz}.Sampling at frequencies above this threshold further separates the alias from $\dez$, keeping the contribution of sampling-induced leakage below the target infidelity.}
    \label{fig:sampling_fft}
\end{figure}

As explained in Section \ref{sec:method}, the fidelity of the resulting operation can be estimated by estimating $S_\text{samp}(-\dez)$. To simplify the expressions, we focus only on the dominant terms $k={0,-1}$ and ignore the remainder. The contribution of a limited sample rate to the infidelity can be calculated by computing the amplitude of the second alias at $-\dez$. 
To keep the infidelity below a certain threshold, the sampling frequency should be chosen as 
\begin{align}
    f_\mathrm{samp}\geq \dez + BW_\mathrm{signal}/2.
 \label{eq:sampling_cri}
\end{align}. 
Here $BW_\mathrm{signal}$ is the bandwidth of the original signal $S(f)$ that must be chosen based on the target infidelity threshold, as shown in the example in Fig.~\ref{fig:sampling_fft}.
However, this analysis does not yet include the effect of the sinc function over the spectrum, which can further reduce the effect of sampling on the infidelity. Additionally, the usage of anti-alias filters or more advanced methods than a pure zero-order hold can further reduce leakage and excitations and improve the gate fidelity.

\begin{figure}[!t]
\centering
\includegraphics[width=.8\columnwidth]{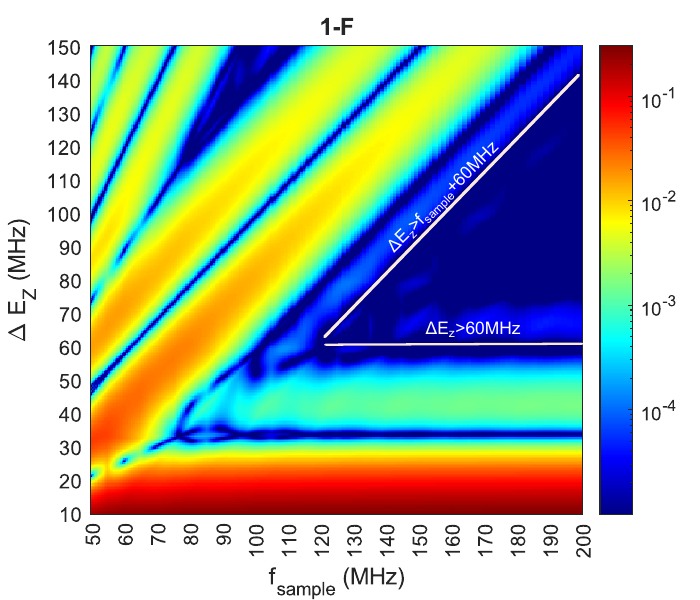}
\caption{ Fidelity of CPHASE operation generated with \unit[50]{ns} raised-cosine signal in two qubit system with different $\dez$ and $f_{\mathrm{sample}}$. In the figure, it can be seen that infidelity is preserved when conditions in Eq.\ref{eq:fourier_error_rate} and Eq.\ref{eq:sampling_cri}. To increase the visibility of the region, the lowest infidelity is limited by $10^{-5}$. The plot with raw data can be found in Fig.~\ref{fig:fig_dez_vs_fsample_sweep_wo_filter}. }
\label{fig:fig_dez_vs_fsample_sweep_w_filter}
\end{figure}

To further investigate the limits in the choice of the sampling frequency, Fig.~\ref{fig:fig_dez_vs_fsample_sweep_w_filter} shows the fidelity of the CPHASE operation generated with a \unit[50]{ns} raised-cosine pulse (which has a $BW_\mathrm{signal}=\unit[120]{MHz}$ for $F_{th}=0.01\%$) for different sampling frequencies and $\dez$. 
High fidelity operations are guaranteed between two boundary conditions: adiabatic leakage (Eq.~\ref{eq:fourier_error_rate}) and minimum sampling frequency (Eq.~\ref{eq:sampling_cri}). 
Outside this region, high-fidelity operations can still be observed under specific conditions. 
Firstly, the $J/\dez$ ratio becomes smaller with large $\dez$, thus, lowers the SWAP oscillation amplitude. 
Secondly, when the sampling period matches the oscillation period, the control signal effectively forms a sequence of diabatic pulses, which can also yield high fidelity. This condition is spectrally equivalent to the sinc envelope in Fig.~\ref{fig:sampling_fft}, exhibiting a notch at $\dez$, thereby suppressing leakage at that frequency.
However, these regimes come with practical limitations. Large $\dez$ can cause strong frequency crowding, thereby constraining addressability and increasing crosstalk in multi-qubit systems. Moreover, the fidelity of diabatic operations is highly sensitive to precise timing and parameter calibration, making them less robust in experimental settings.

\section{Discussion}
\label{sec:conclusion}
This work introduces the time-domain pulse-shaping strategy Delayed Leakage Reduction (DLR), that enables high-fidelity, ultra-fast adiabatic operations in systems constrained to baseband control, such as two-qubit gates for spin qubits. By analytically linking the gate fidelity to the spectral properties of the control signal, we reformulate the fidelity optimization problem as a spectral design problem. The proposed DLR technique introduces targeted spectral notches through time-delayed pulse replicas, effectively suppressing leakage at critical transition frequencies without requiring IQ modulation. Simulation results demonstrate that DLR achieves fidelity exceeding 99.9\% within gate durations as short as \unit[9.4]{ns}, while only marginally increasing the hardware sampling requirements.

While DLR is primarily designed to suppress leakage at a single transition frequency, it can be extended to multi-body systems by applying it sequentially. In this approach, each DLR application is tuned to place a spectral notch at a different target frequency. However, since each additional DLR pulse shortens the effective gate duration, the spectrum of the resulting signal broadens, and previously placed notches may shift. This shift can be compensated by ensuring that the product of the notch frequency and the gate duration remains constant, i.e., $f_{\mathrm{notch},1} \cdot t_{\mathrm{gate},1} = f_{\mathrm{notch},2} \cdot t_{\mathrm{gate},2}$. This relation allows earlier DLR stages to be adjusted accordingly, preserving the intended spectral suppression across multiple transitions.

Thanks to its conceptual generality and compatibility with baseband control, the proposed DLR technique is not only applicable to spin qubits but can also be extended to other qubit platforms and adiabatic processes. By enabling ultra-fast, high-fidelity gate operations that surpass the speed of current state-of-the-art two-qubit spin-qubit gates, DLR offers a versatile and scalable tool for advancing the development of next-generation quantum processors.

%Code Availability
\section*{Code Availability}
The source code used for pulse optimization and fidelity simulations is available at \url{https://doi.org/10.5281/zenodo.16707554} \cite{polat_2025_16707554} . The repository includes scripts for reproducing all figures and numerical results presented in this work.

%Acknowledgments
\section*{Acknowledgments}

This research is partially supported by Intel and partly sponsored by the Army Research Office under Award Number: W911NF-23-1-0110. 
The views and conclusions contained in this document are those of the authors and should not be interpreted as representing the official policies, either expressed or implied, of the Army Research Office or the U.S. Government. 
The U.S. Government is authorized to reproduce and distribute reprints for Government purposes notwithstanding any copyright notation herein.

%Appendix

\appendix

\appendix
\section{Window Functions for Spectral Shaping}
\label{app:window_functions}

In this appendix, we summarize the mathematical definitions of the three window functions employed for shaping the exchange control envelope $J(t)$, Raised-Cosine, Kaiser, and Hamming windows. These functions are widely used in signal processing to suppress spectral leakage and to tailor the energy distribution of time-domain control signals.

Each window function is defined over the interval $t \in [0, t_\text{gate}]$, where $t_\text{gate}$ denotes the total gate duration. Outside this interval, i.e., for $t < 0$ or $t > t_\text{gate}$, the window function is identically zero:
\begin{equation}
w(t) = 0, \quad \text{for } t < 0 \text{ or } t > t_\text{gate}.
\end{equation}

\subsection{Raised-Cosine Window}
The Raised-Cosine window is defined as
\begin{equation}
w_{\mathrm{RC}}(t) = \frac{1}{2} \left[1 - \cos\left(2\pi \frac{t}{t_\text{gate}}\right)\right],
\end{equation}
for $t \in [0, t_\text{gate}]$. This window exhibits smooth onset and termination, effectively suppressing high-frequency components in the spectral domain.

\subsection{Kaiser Window}
The Kaiser window introduces a tunable shape parameter $\beta$ that controls the trade-off between main-lobe width and side-lobe suppression. It is defined as
\begin{equation}
w_{\mathrm{K}}(t) = \frac{I_0\left(\pi \beta \sqrt{1 - \left(\frac{2t}{t_\text{gate}} - 1\right)^2}\right)}{I_0(\pi \beta)},
\end{equation}
where $I_0(\cdot)$ denotes the zeroth-order modified Bessel function of the first kind. Larger values of $\beta$ yield stronger side-lobe suppression at the cost of broader main lobes.

\subsection{Hamming Window}
The Hamming window is a fixed-shape window function defined as
\begin{equation}
w_{\mathrm{H}}(t) = 0.54 - 0.46 \cos\left(2\pi \frac{t}{t_\text{gate}}\right),
\end{equation}
for $t \in [0, t_\text{gate}]$. Compared to the Raised-Cosine window, the Hamming window offers an improved main-lobe width at the expense of slightly higher side-lobe levels.

All window functions are normalized such that
\begin{equation}
\int_0^{t_\text{gate}} w(t) \, dt = t_\text{gate},
\end{equation}
ensuring consistent energy scaling across different pulse shapes.

\section{Simulations}
\label{appx:simu}

Simulation scripts are developed using the SPINE \cite{DijkSPINE} toolkit. To solve the time-dependent evolution of the qubit system, the Schrödinger equation

\begin{align}
	i  \hbar \frac{d}{dt}|\psi(t)\rangle = H(|\psi(t)\rangle)
\end{align}
is solved iteratively as
\begin{align}
	U(t+\Delta t)=e^{- \frac{i }{\hbar}H(t+\Delta t)\Delta t}U(t),
\end{align}
 where the quantum operator $U$ is calculated with small time steps $\Delta  t$ by assuming that the change of $H$ between steps is negligible. The average gate fidelity is given by
\begin{align}
F\left ( U_{\mathrm{ideal}}, U_{\mathrm{sim}}\right) &= \frac{ {\mathrm {Tr}} \left[\mathcal U_{\mathrm{ideal}}^\dagger \mathcal  U_{\mathrm{sim}}\right] + d} {\left(d+1 \right)d}\\ &= \frac{ \text{Tr}\left|U^\dagger_{\mathrm{ideal}}U_{\mathrm{sim}}\right|^2 + d} {\left(d+1 \right)d},
\label{eq:averageGateFidelity}
\end{align}
where the superoperators are calculated as
\begin{align}
	\mathcal U = U^*\otimes U.
\end{align}

\section{Explicit expressions of errors}
\label{app:errors}
Following ~\cite{Rimbach-Russ_2023}, the total imperfect time evolution of an adiabatic quantum gate can be written as
\begin{align}
    U_\text{gate}(t) =  \mathcal{D}^\dagger(t)U_\text{ad}(t)\mathcal{E}(t)\mathcal{D}(0)
\end{align}
for all times $t$ in the interval $[0,t_\text{gate}]$. Here, $U_\text{ad}$ is the desired operation to perform, $\mathcal{E}$ is the error, and $\mathcal{D}$ is the unitary transformation from the computational basis into the instantaneous eigenbasis. Remark that for our case $\mathcal{D}(0)=\mathcal{D}(t_\text{gate})=1$, thus we will ignore the terms in the following. For adiabatic dynamics 
\begin{align}
    U_\text{ad}(t) &= \exp\left[-\frac{i  }{\hbar} \int_0^{t} H_\text{ad}(t^\prime) dt^\prime\right]\\
    &= \sum_{m} e^{i\alpha_m(t)} \ket{m(t)}\bra{m(t)}.
\end{align}
where we used the spectral decomposition, introduced the phases $\alpha_m = - \frac{1}{\hbar} \int_{0}^{t} \epsilon_m(t^\prime)dt^\prime$, eigenvectors $\ket{m}$, and eigenenergies $\epsilon_m(t)$. The error term can be expressed as a Magnus expansion~\cite{Rimbach-Russ_2023}
\begin{align}
    \mathcal{E}(t) = \exp\left[ -\frac{i }{\hbar} \sum_{k=1}^\infty H_\text{Mk} \right],
    \label{app:magnusseries}
\end{align}
where the lowest order term is given by integration over the error Hamiltonian
\begin{align}
    H_\text{M1} = \frac{1}{\hbar} \int_0^t H_\text{error}(t^\prime)dt^\prime.
    \label{app:magnus1}
\end{align}
The error Hamiltonian is given by~\cite{Rimbach-Russ_2023}
\begin{align}
    H_\text{error} = \sum_{m_1,m_2}e^{i  (\alpha_{m_2}-\alpha_{m_1})}\bra{m_1(t)}H_\text{total}-H_\text{ad}\ket{m_2(t)}
\end{align}
and finally the error rates for the transition $\mathcal{O}_{m_1,m_2}=\ket{m_2}\bra{m_1}$ by

\begin{align}
    \epsilon_{m_1,m_2}=&|\text{Tr}\left(\mathcal{\mathcal{E}}\mathcal{O}_{m_1,m_2}-\mathcal{O}_{m_1,m_2}\right)|^2\\
    \approx & \left|\frac{1}{\hbar} \int_{0}^{t_\text{gate}} \bra{m_1(t)}H_\text{total}(t)\right.\nonumber\\
    &-\left. H_\text{ad}(t)\ket{m_2(t)} e^{i  (\alpha_{m_2}(t)-\alpha_{m_1}(t))}  \right|^2 
    \label{app:errorrates}
\end{align}

For the coupling between two spin qubits, the Hamiltonians (in frequency units) read
\begin{align}
    H_\text{ad}=&\frac{h}{2}\begin{pmatrix}
    2Ez & 0         & 0         & 0 \\
    0   & f_\text{osc} -J     & 0         & 0 \\
    0   & 0       & -f_\text{osc} -J      & 0 \\
    0   & 0         & 0         & -2E_z
\end{pmatrix}
\end{align}
\begin{align}
H_\text{total}-H_\text{ad} = &\frac{h}{2}\begin{pmatrix}
    0 & 0         & 0         & 0 \\
    0   & 0     & \frac{i }{2\pi}\frac{\dot{J}(t)\Delta E_z}{f^2_\text{osc}}         & 0 \\
    0   & -\frac{i }{2\pi}\frac{\dot{J}(t)\Delta E_z}{f^2_\text{osc}}       & 0      & 0 \\
    0   & 0         & 0         & 0
\end{pmatrix},
\end{align}
where $f_\text{osc}=\sqrt{J^2+\Delta E_z^2}$ is the eigenfrequency and $\dot{J}$ the time derivative of the exchange frequency. The single non-adiabatic error is then given by~\cite{Rimbach-Russ_2023}
\begin{align}
    \epsilon = |\text{Tr}(\mathcal{\mathcal{E}}\ket{01}\bra{10})|^2 = \left|\frac{\Delta E_z}{2}\int_{0}^{t_\text{gate}} \frac{\dot{J}}{f_\text{osc}^2}e^{-2\pi i  \int_0^t \fosc(t^\prime)dt^\prime} \right|^2.
\end{align}
The expression from the main text, Eq.~\eqref{eq:sparseEvolution} can be recovered by using integration by parts assuming $J(0)=J(t_\text{gate})=0$
\allowdisplaybreaks
\begin{align}
    \epsilon =& \left|\pi\int_{0}^{t_\text{gate}}\fosc(t) \arctan\left(\frac{J(t)}{\Delta E_z}\right)e^{-2i\pi  \int_0^t \fosc(t^\prime)dt^\prime} \right|^2\\
    \approx& \left|\pi\int_{0}^{t_\text{gate}}J(t)e^{-2i\pi  \int_0^t \fosc(t^\prime)dt^\prime} \right|^2.
\end{align}
In the last line, we used $\fosc(t)\arctan\left(\frac{J(t)}{\Delta E_z}\right)= J(t) +\mathcal{O}\left(\frac{J(t)^3}{\Delta E_z^2}\right) $.

\section{Second-order Magnus expansion}
\label{app:magnus}
The correction from second-order Magnus expansion is given by Eq.\eqref{eq:magnus2} where the explicit expressions read
\begin{align}
\delta &= 4\pi^2\text{Im}\left[\int_0^{t_\text{gate}}dt\int_{0}^t dt^\prime g(t)g^\star(t^\prime)e^{-i 2\pi (\alpha(t)-\alpha(t^\prime))} \right]
\end{align}
\begin{align}
    g(\tau)&= -\frac{\dez\dot{J}(\tau)}{2\pi \fosc^2(\tau)}
\end{align}
\begin{align}
    \alpha(\tau)&=\int_0^{\tau}dt^{\prime\prime}\fosc(t^{\prime\prime}).
\end{align}

%\begin{widetext}
\begin{align}
    H_\text{M2} &=\frac{i }{2\hbar^2}\int_0^{t_\text{gate}}dt\int_{0}^t dt^\prime \left[H_\text{error}(t),H_\text{error}(t^\prime)\right]\\
          &=i 2\pi^2\int_0^{t_\text{gate}}dt\int_{0}^t dt^\prime \bigg( g(t)g^\star(t^\prime)e^{-i 2\pi (\alpha(t)-\alpha(t^\prime))}\nonumber\\
          &-g(t^\prime)g^\star(t)e^{i 2\pi (\alpha(t)-\alpha(t^\prime))} \bigg)\begin{pmatrix}
    0 & 0         & 0         & 0 \\
    0   & 1     & 0         & 0 \\
    0   & 0       & -1      & 0 \\
    0   & 0         & 0         & 0
\end{pmatrix}\\
          &=\begin{pmatrix}
    0 & 0         & 0         & 0 \\
    0   & \delta     & 0         & 0 \\
    0   & 0       & -\delta      & 0 \\
    0   & 0         & 0         & 0
\end{pmatrix},
\label{eq:magnus2}
\end{align}

%\end{widetext}

\section{Phase calibration, final gate time, and average gate fidelity}
\label{appx:relation_fidelity_and_spectrum}

%\begin{widetext}

%\end{widetext}
\begin{widetext}
The average gate fidelity of the total gate evolution $U_\text{gate}(t_\text{gate})$ with respect to the targeted quantum gate $U_\text{ideal}=Z_1(-\theta_1)Z_2(-\theta_2)\text{CZ}$ is then given by 

\begin{align}
    F(U_\text{ideal},U_\text{gate})=F\left[U_\text{ideal},\mathcal{D}^\dagger(t_\text{gate})U_\text{ad}(t_\text{gate})\mathcal{E}\mathcal{D}(0)\right],\\
    \approx F\left[U_\text{ideal},U_\text{ad}(t_\text{gate})\exp\left(-\frac{i }{\hbar}(H_\text{M1}+H_\text{M2})\right)\right],\\
    = F\left[\text{CZ},Z_1(\theta_1)Z_2(\theta_2)U_\text{ad}(t_\text{gate})\exp\left(-\frac{i }{\hbar}(H_\text{M1}+H_\text{M2})\right)\right],
    \label{eq:aux1}
\end{align}
where we used the approximation and simplifications made in App.~\ref{app:errors} and the properties of the trace. We note that the two single-qubit phase gates $Z_1(-\theta_1)$ and $Z_2(-\theta_2)$ can be done digitally by using the virtual gate method. Using further $\text{CZ}=\text{CZ}^\dagger$, the fidelity can be expressed as

\begin{align}
    F(U_\text{ideal},U_\text{gate})&=F\left[1,\exp\left(-i  \Phi_J Z_1 Z_2\right)\exp\left(-i  \delta\Phi (Z_1+Z_2)/2\right)\exp\left(-i  \delta\theta (Z_1-Z_2)/2\right)\exp\left(-\frac{i }{\hbar}(H_\text{M1}+H_\text{M2})\right)\right],
    \label{eq:aux2}
\end{align}
\end{widetext}
where $\delta\Phi_J=\pi-2\pi\int_0^{t_\text{gate}}J(t)dt$ is the exchange miscalibration, $\delta\Phi= (\theta_1 + \theta_2-\pi)-2\pi E_z t_\text{gate}$ the miscalibration of the rotating frame, and $\delta\theta = (\theta_1 - \theta_2)-2\pi\int_0^{t_\text{gate}}\sqrt{\Delta E_z^2+ J^2(t)}dt$ the miscalibrated phase corrections. Assuming small errors and using the explicit error Hamiltonians~\eqref{app:magnus1} and \eqref{eq:magnus2}, we find the following compact expressions for the infidelity by expanding Eq.~\eqref{eq:aux2} up to second order
\begin{align}
    1-F(U_\text{ideal},U_\text{gate})&\approx \frac{2}{5}\left[\epsilon+2\Phi_j^2+(\delta\theta+\delta)^2+\delta\Phi^2\right].
\end{align}
Setting $\delta=0$, we recover formula~\eqref{eq:fidelityAnalytic} from the main text.

\begin{figure}[t]
\centering
\includegraphics[width=\columnwidth]{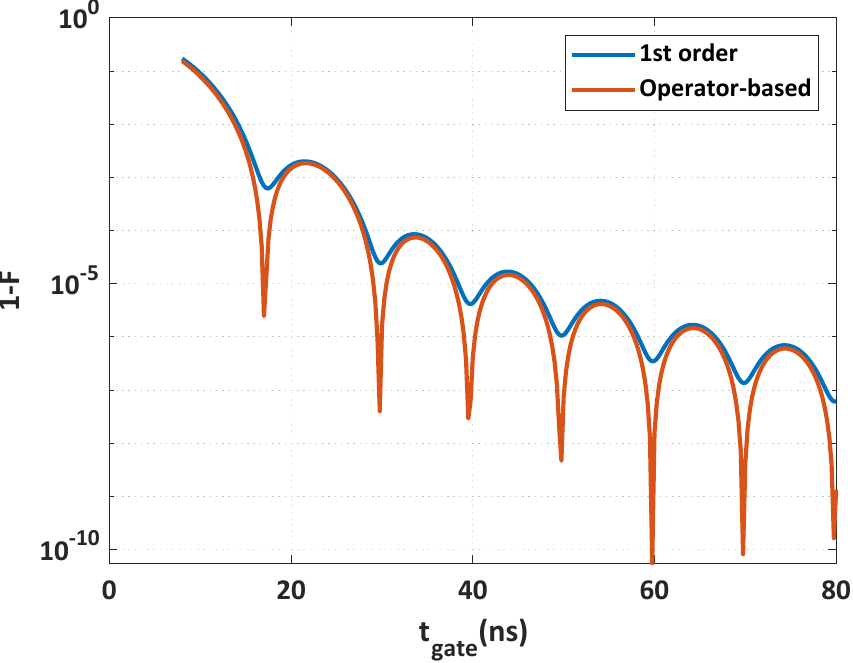}
\caption{   Infidelity (1-F) of the CPHASE gate plotted against gate time $t_{\mathrm{gate}}$, comparing the 1st--order
            approximation (blue) and the operator--based approach (orange).}
\label{fig:z_calibration}
\end{figure}

\section{Numerical Phase Calibration}
\label{appx:z_calibration}

While $\int^{t_\text{gate}}_0 J(t)$ generates phase accumulation of $\pi$, this operation also shifts the oscillation frequencies of the qubits during the operation. This can be calibrated by a single qubit $Z_{1,2}$ explained in Appen.\ref{appx:relation_fidelity_and_spectrum}. 
The rotation angle of this calibration can be calculated as 
\begin{align}
        \theta_{1,2}  = \dez \mp \sqrt{\dez^2+J^2}. 
	\label{eq:z_calibration_1st}
\end{align}

Then, $Z_1(\theta_1)$ and $Z_2(\theta_2)$ can be applied to qubits. However, during the simulations, we realized that Eq.\eqref{eq:z_calibration_1st} doesn't give the best results. To improve the accuracy, a higher-order Magnus expansion can be used.

\begin{figure*}[!ht]
    \subcaptionbox{\label{fig:DLR_static_sweep_tukey}}{\includegraphics[width=0.3\textwidth]{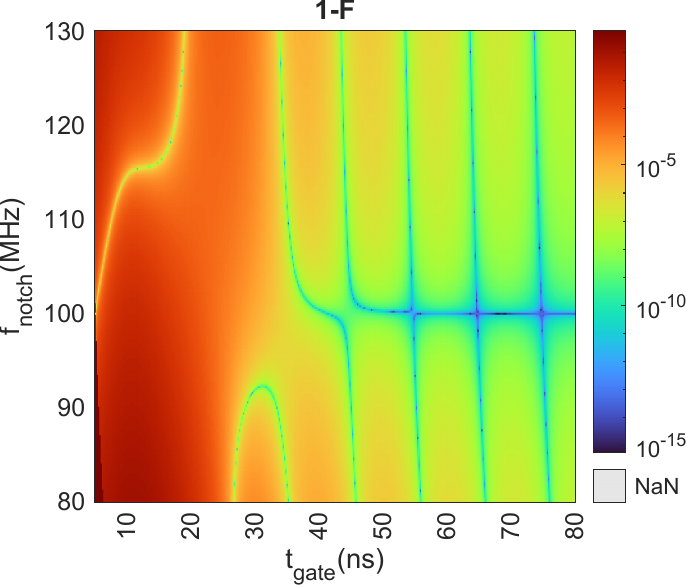}}
    \hfill
    \subcaptionbox{\label{fig:DLR_static_sweep_hamming}}{\includegraphics[width=0.3\textwidth]{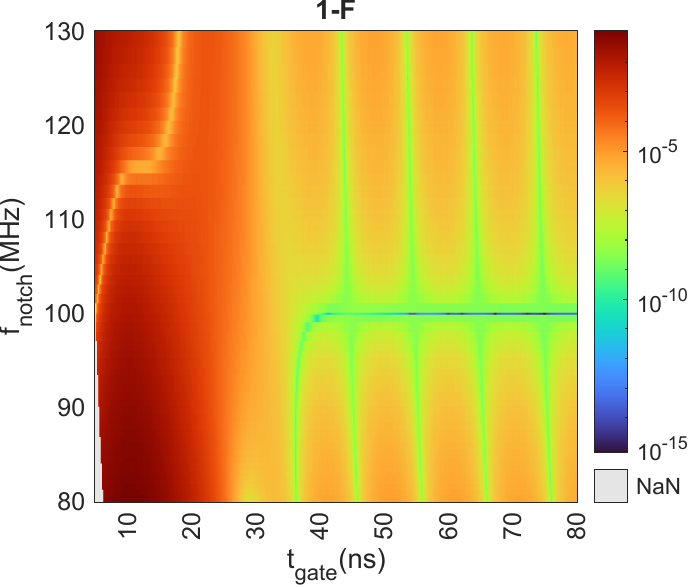}}
    \hfill
    \subcaptionbox{\label{fig:DLR_static_sweep_kaiser}}{\includegraphics[width=0.3\textwidth]{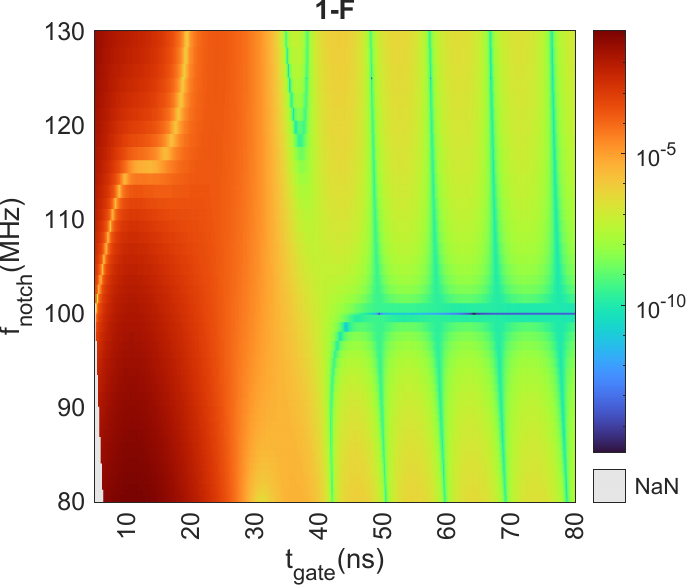}}
    \vfill
    \subcaptionbox{\label{fig:DLR_static_sweep_tukey_contour}}{\includegraphics[width=0.3\textwidth]{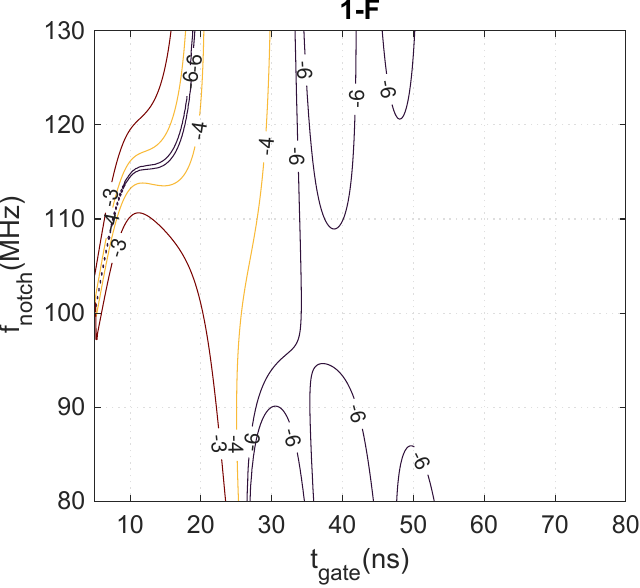}}
    \hfill
    \subcaptionbox{\label{fig:DLR_static_sweep_hamming_contour}}{\includegraphics[width=0.3\textwidth]{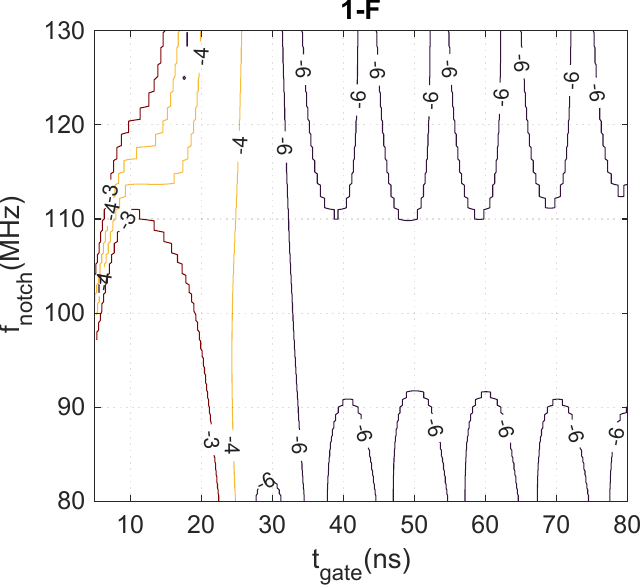}}
    \hfill
    \subcaptionbox{\label{fig:DLR_static_sweep_kaiser_contour}}{\includegraphics[width=0.3\textwidth]{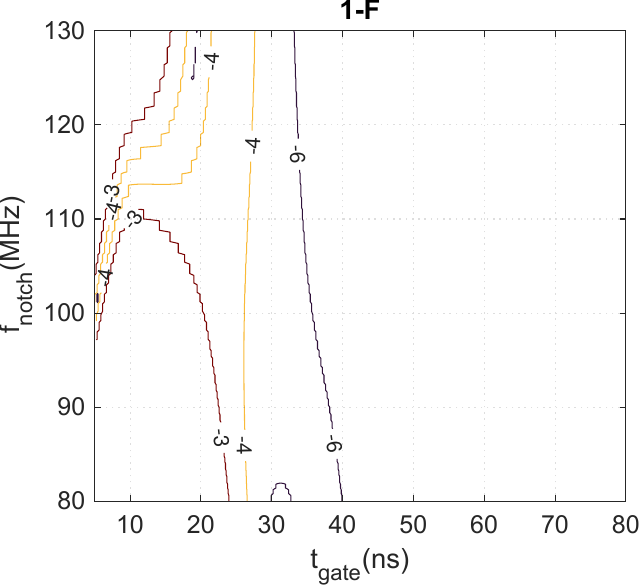}}
    
    \caption{ The notch created on the frequency spectrum of the signal can be chosen arbitrarily. To calibrate the DLR numerically, a range of notch frequencies is tested for different gate times. Here, infidelities are shown for (a) raised-cosine, (b) Hamming, and (c)  Kaiser windows. The corresponding fidelity boundaries at 0.1\%, 0.01\%, and 0.0001\% are shown in panels (d–f), providing a visual reference for the performance thresholds across the parameter space.}
    \label{fig:DLR_static}
\end{figure*}

While numerically solving these equations may not be the most straightforward method, $\theta$ is derived from the evolution operator. The unitary propagator in a rotating reference frame, $U$, is determined as described in App.~\ref{appx:simu}. Subsequently, the angles $\theta_{1,2}$ are calculated as $\theta_1 = \pi + \angle{U_{11}} - \angle{U_{22}}$ and $\theta_2 = \pi + \angle{U_{44}} - \angle{U_{33}}$, where $U_\text{row,column}$ denotes the matrix elements of $U$. Consequently, the difference in angle between the states $|0\rangle$ and $|1\rangle$ for a single qubit is $\pi$.

In Fig.~\ref{fig:z_calibration}, comparison of Eq.~\eqref{eq:z_calibration_1st} and operator-based method. It can be seen that the accuracy of the CPHASE is improved, especially on an integer multiple of the SWAP--oscillation periods ($\approx$\unit[10]{ns}). 

\section{Comparison with DRAG}
In the derivative removal by adiabatic gate (DRAG) protocol, the spectral nodes are constructed using I/Q control. We emulate this here by considering complex exchange signals $J$. In particular, we modify the exchange Hamiltonian in Eq.~\eqref{eq:hexec} with
\begin{align}
H_{exc'} = \begin{pmatrix}
    E_z & 0         & 0         & 0 \\
    0   & (-\dez-|J|)/2     & \Bar{J}/2         & 0 \\
    0   & J/2       & (\dez-|J|)/2      & 0 \\
    0   & 0         & 0         & -E_z
\end{pmatrix}
\label{eq:hexec_new}
\end{align}

\section{Additional Data of DLR Numerical Optimization}
\label{appx:dlr_numerical}

Due to the initial discontinuity present in the Hamming and Kaiser window functions, Eq.~\eqref{eq:fourier_error_rate} no longer holds. Consequently, we resort to a direct numerical optimization approach, where the delay parameter $t_d$ in the DLR protocol is swept between \unit[3.84]{ns} and \unit[6.25]{ns}, corresponding to a notch frequency $f_{\mathrm{notch}}$ ranging from \unit[130]{MHz} to \unit[80]{MHz}, respectively. The resulting gate fidelities for each combination of gate time and $f_{\mathrm{notch}}$ are shown in Fig.~\ref{fig:DLR_static}.

For all window functions considered, longer gate durations consistently yield optimal performance when $f_{\mathrm{notch}} \approx \dez \approx \unit[100]{MHz}$. The vertical stripe patterns observed in the fidelity maps arise from non-adiabaticity conditions, where residual oscillations synchronize with the gate duration. For shorter gate times, however, the assumption $f_{\mathrm{osc}} \approx \dez$ breaks down, and the optimal choice of $f_{\mathrm{notch}}$ becomes window-function dependent.

Beyond \unit[35]{ns}, as $f_{\mathrm{osc}}$ increases, higher values of $f_{\mathrm{notch}}$ tend to yield improved performance. However, at sufficiently large $f_{\mathrm{osc}}$ at the lower gate time, the main spectral lobe of the pulse begins to overlap with $\dez$, leading to increased leakage. In this regime, selecting a lower $f_{\mathrm{notch}}$ becomes advantageous. The optimal fidelity achieved for each gate duration is summarized in Fig.~\ref{fig:fig_DLR_static_sweep}.

\section{Additional Data of Sampling Frequency}
\label{appx:sampling_frequency}

Simulated fidelity of the $\mathrm{CPHASE}$ operation generated by a \unit[50]{ns} raised-cosine pulse in a two-qubit system, as a function of the qubit frequency separation $\dez$ and the sampling frequency $f_{\mathrm{sample}}$. The data shown is identical to Fig.~\ref{fig:fig_dez_vs_fsample_sweep_w_filter}; however, no clipping filter has been applied to the fidelity values. In Fig.~\ref{fig:fig_dez_vs_fsample_sweep_w_filter}, fidelity values below $10^{-5}$ were clipped for improved visibility, while here the raw fidelity values are retained. This allows for a more accurate representation of the fidelity landscape, particularly in regions where the error rate falls below the clipping threshold.

\begin{figure}[!htbp]%tbp
\centering
\includegraphics[width=\columnwidth]
{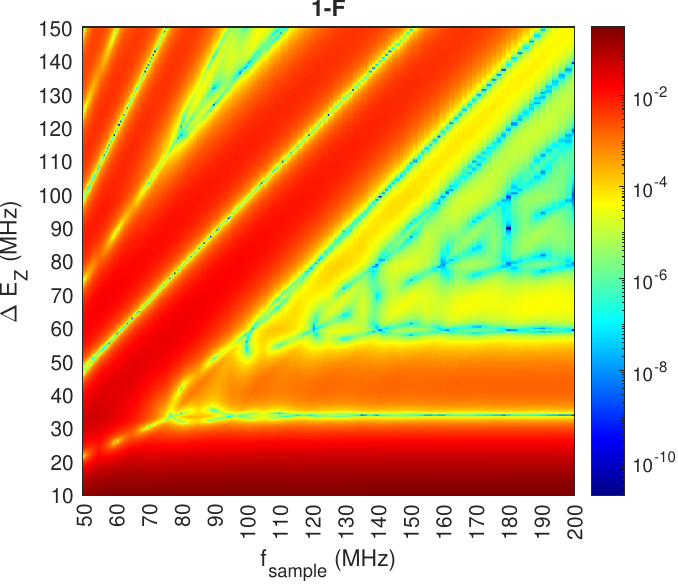}
\caption{ Fidelity of CPHASE operation generated with \unit[50]{ns}  Raised-Cosine signal in two qubit system with different $\dez$ and $f_{\mathrm{sample}}$. }
\label{fig:fig_dez_vs_fsample_sweep_wo_filter}
\end{figure}

\FloatBarrier

\bibliography{draftNotes}

\end{document}